\documentclass[usenatbib]{mn2e}
\usepackage {graphicx}
\begin{document}
\title[Globular Clusters as Tracers of Stellar Bi-modality in Elliptical
Galaxies]{Globular Clusters as Tracers of Stellar Bi-modality in Elliptical
Galaxies: The Case of NGC 1399}
\author[Juan C. Forte, Favio Faifer \& Doug Geisler ]{Juan C. Forte $^{1,2}$\thanks{E-mail:forte@fcaglp.unlp.edu.ar},
Favio Faifer$^{1,2}$ and Doug Geisler$^{3}$\\
$^1${Facultad de Ciencias Astron\'omicas y Geof\'{\i}sicas, Universidad Nacional de La Plata}\\
$^2${Consejo Nacional de Investigaciones Cient\'{\i}ficas y T\'ecnicas, Rep.
Argentina}\\
$^3${Departamento de F\'{\i}sica, Universidad de Concepci\'on, Chile}\\}
\date{\today}
\maketitle
\label{firstpage}

\begin{abstract}
Globular cluster systems (GCS) frequently show a bi-modal distribution of
the cluster integrated colours. This work explores the arguments to support
the idea that the same feature is shared by the diffuse stellar
population of the galaxy they are associated with. The particular
case of NGC 1399, one of the dominant central galaxies in the Fornax
cluster, for which a new $B$ surface brightness profile and
$(B-R_{kc})$ colours are
presented, is discussed taking advantage of a recently published wide
field study of its GCS. The results show that the galaxy 
brightness profile and colour gradient as well as the behaviour of the
cumulative globular  cluster specific frequency, are compatible with the
presence of two dominant stellar populations, associated with
the so called ``blue'' and ``red'' globular cluster families. These
globular families are characterized by different intrinsic specific
frequencies (defined in terms of each stellar population): $Sn=3.3 \pm 0.3$
in the case of the red globulars and $Sn=14.3 \pm 2.5$ for the
blue ones. We stress that this result is not necessarily conflicting
with recent works that point out a clear difference between the
metallicity distribution of (resolved) halo stars and globulars
when comparing their number statistics.
The region within 0.5 arcmin of the centre shows a deviation
from the model profile (both in surface brightness and colour) that
may be explained in terms of the presence of a bulge-like high
metallicity component. Otherwise, the model gives an excellent fit
up to 12 arcmin (or 66.5 Kpc) from the centre,
the galactocentric limit of our blue brightness
profile. The inferred specific frequencies imply that, in terms of their
associated stellar populations, the formation of the blue globulars took
place with an efficiency about 6 times higher than that corresponding to
their red counterparts.
The similarity of the spatial distribution of the blue globulars with
that inferred for dark matter, as well as with that of the X ray emiting
hot gas associated with NGC 1399, is emphasized. 
The impact of a relatively unconspicuous low metallicity
population, that shares the properties of the blue globulars,
as a possible source of chemical enrichment early in the 
formation history of the galaxy is also briefly discussed.
\end{abstract}


\section{Introduction}

A large volume of arguments supports the idea that the current structure of most galaxies has been determined by events that occurred at the very early epochs of the Universe (see, for example, \citealt{BAUGH}, and references therein) . In this context, and being the oldest stellar systems for which a reliable chronology exists (at least in the Milky Way), globular clusters (GC) have become the subject of an increasing amount of research that aims at identifying the link between their formation and that of their host galaxies themselves.\\
\indent Even though the fact that GC may be used as probes of the formation of a galaxy has been known for a long time (e.g. \citealt{ELS62}) and that the knowledge of globular cluster systems (GCS), in our and in other galaxies, has been steadily growing, a unified picture that explains both the small scale phenomena that lead to the cluster formation, and also their overall properties as systems, is still missing.\\
\indent Recent reviews give good summaries of the situation (e.g. \citealt{Harris3}; \citealt{KP2002}; \citealt{KP2003}) and compare the features of several astrophysical scenarios that invoke ``external'' or ``internal'' phenomena that provide a reasonable confrontation with observations. It seems clear, however, that some aspects of the problem require a cosmological frame as already suggested by \citet{PD68}. More recently, other attempts assume hierarchical CDM models \citep{B2002} that appear as a promissing approach to further incorporate the effects of different and co-existing (rather than excluding) processes that end up with the formation of a GCS as, for example, the role of GC as a possible source of the re-ionization of the Universe (\citealt{Rico2002}; \citealt{San2003}).

All those scenarios, however, must take into account some relatively well established facts as boundary conditions, namely:

\begin{itemize}
\item[-]Stellar clusters that fulfill the properties expected for a ``young''
globular cluster are currently formed in violent events, such as galaxy
mergers, as originally pointed out by \citet{Schw87}. These kind
of phenomena, on a variety of scales, were probably more frequent
during the early phases of galaxy formation.

\item[-]GC formation seems connected with major stellar formation events.
This fact is well supported by the relatively narrow range of
the ratio of the number of GC to galaxy luminosity, as well as by the
correlation of cluster formation with high star formation rate \citep{LR2000}.
This ratio can be quantified, for example, using the so called 
``specific globular cluster frequency'', Sn, following \citet{HB81}.
We point out that a revision of several archetypical high Sn galaxies
have lowered previous estimates of this parameter by half, or even less.
(e.g. \citealt*{OFG98}; \citealt{Forte2002}).

\item[-]The Milky Way  GC
metallicity distribution shows a bi-modal structure. This feature seems
shared by other galaxies that show bi-modal colour distributions
(e.g. \citealt*{GLK96}). Although the well known colour-age-metallicity 
degeneracy prevents a stronger statement about the significance of the
colour distribution, the assumption of very old ages for GC points
to metallicity as the main driver of the colour distribution. 

\item[-]In the case of NGC 1399, the so called ``red'' GC exhibit
a distinct behavior, when compared with the ``blue'' GC, both in terms
of their spatial distribution and kinematic behaviour (e.g. \citealt{R2004}; R2004 in what follows).
\end{itemize}

In this picture, the connection between the GCS and the stellar
population in a galaxy is not completely clear. Some differences
\citep*{FSS81} and some similarities \citep{FF2001}
have been pointed out over  the years. This last paper, in
particular, reinforces previous arguments \citep{FBH97}
that suggest a connection between the red GC and the galaxy
halo stars based on the similarity of the red GC colours
with those observed in the {\textit {inner}} regions of a galaxy.
A significant contribution to the subject has been presented
by \citet{HH2002} (and references therein) who discuss the
characteristics of the resolved stellar population of NGC 5128
and present a comparison with the GC in that galaxy. These
authors also discuss a possible connection between GC specific
frequency and metallicity.

This paper explores the arguments to support (or reject) the idea that,
in the bi-modal GCS, the ``diffuse'' galaxy stellar population also exhibits
a dual nature reflecting the existence of two major stellar
populations that share the properties of each GC family and are
typified by a given ``intrinsic'' Sn, integrated colour, and spatial
distribution. 

We emphasize that this work presents an approach that, in principle,
makes no assumptions about the nature of integrated colours (in terms
of ages or metallicities) and finally aims at matching the shape of the
observed galaxy surface brightness, colour gradient and cumulative Sn
(with galactocentric radius) by properly weighting each of the
diffuse stellar components. These stellar populations, defined just
in colour terms, then might include a range of age and/or metallicity
as well as of GC specific frequencies.

NGC 1399, the central galaxy in the Fornax cluster, appears as a
good candidate on which to perform this kind of analysis since, as in other
luminous ellipticals, its GCS is well populated and can provide
statistically significant results. In addition, it is relatively nearby.

A number of papers have dealt with the clearly bi-modal GC colour distribution
in this galaxy taking advantage of the high metallicity sensitivity of the 
colour index defined in terms of the $C$ and $T_1$ bands of the Washington 
photometric system \citep{C76} . These early 
works (\citealt{GF90}; \citealt{OGF93}; \citealt{OFG98}) have a relatively 
small areal coverage. In turn, \citet*{D2003} (D2003 hereafter) increased 
the surveyed area to about 900  sq. arcmin by means of mosaic CCD observations 
and determined GC areal density profiles up to a galactocentric radius of 
about 20 arcmin. These density profiles, and the surface photometry
(Johnson's $B$ and Kron-Cousins $R$ filters) presented 
in this work are the basis of the subsequent analysis. An attempt 
to disentangle the stellar populations in NGC 1399 was presented in D2003
although, as described later, with a somewhat different approach.

The structure of the paper is a follows: Section~\ref{CSBP} describes the
result of composing two stellar populations and the variables that
govern the shape of the galaxy surface brightness profile as well
as the derivation of the intrinsic specific frequencies of
each GC population; The characteristics of the GCS system associated
with NGC 1399 and a justification of the adopted distance modulus are
presented in Section~\ref{GCSD}; a new blue profile and $(B-R_{kc})$ colors  
for the galaxy are presented in Section~\ref{BandR1399};
The areal density run with galactocentric radius for both the
red and blue GC, as well as the derivation of
their volumetric density profiles, are discussed in
Sections~\ref{ADP} and~\ref{VDP}, respectively; Section~\ref{RES}
gives a confrontation of the observations with
the properties of the model in terms of the shape of the surface
brightness profile, colour gradient and cumulative GC specific
frequencies as a function of  galactocentric radius. 
A caveat about the approach is commented  on in 
Section~\ref{UNCER}. A first approach to the expected metallicity
distribution function of the galaxy halo stars is described
in Section~\ref{SMDF}; The total stellar mass as well as the GC
formation efficiencies are discussed in Section~\ref{SMEFF}; A comparison
of the GC spatial distribution with that inferred for dark matter
(R2004) and
hot gas obtained from X ray observations are given in Section~\ref{COMP}; A
discussion of the results and possible implications on the galaxy
formation scenario are included in Sections~\ref{DISC} and~\ref{CONCL} 
respectively.

\section{A Composite Surface Brightness Profile}
\label{CSBP}

In what follows we assume that both the red and blue globulars
in NGC 1399 have an associated diffuse stellar population
characterized by their own intrinsic specific frequencies.

A generalisation of the $Sn$ parameter to any photometric band,
and in particular, for the $B$ (Johnson's blue) band, leads to a
composite luminosity (in $M(B)=-15$ units):

\begin{equation}
L(B) = \frac{N(RGC)}{S_B(RGC)}+\frac{N(BGC)}{S_B(BGC)}
\end{equation}

The blue band is chosen, in this case, since we will attempt to
match the shape of the $B$ surface brightness profile presented in
Section 4. If $N(RGC)$ and $N(BGC)$ are the projected areal
densities of the red and blue GC respectively and $(V-Mv)_o$ is the
(interstellar extinction) corrected distance modulus, for any
galactocentric radius:

\begin{equation}
(V-Mv)_o= \mu_B-A(B)-M(B)
\end{equation}

\noindent where $\mu_B$ is the galaxy surface brigntness, $A(B)$ is the 
interstellar extinction along the line of sight in the blue band, and
$M(B)=-2.5log(L(B)\times10^6)$ is the areal absolute blue magnitude,
if $L(B)$ is given in absolute brightness units of $M(B)=-15.0$; and if
both intrinsic specific frequencies are assumed to remain
constant with galactocentric radius, then it follows:
\begin{eqnarray}
\mu_B= (V-Mv)_o+A(B)+2.5log\big[ S_B(RGC) \big] \nonumber  \\
-2.5log \big[ N(RGC)+\frac{N(BGC)}{C(B)} \big]
\end{eqnarray}

\noindent where, $C(B)$ is the ratio of the specific frequencies of the 
blue to red GC populations, $S_B(BGC)/S_B(RGC)$.

The expression for $\mu_B$ shows that, if the areal densities of
both cluster families as a function of galactocentric radius
are determined from observations, the shape of the surface
brightness profile will depend only on the $C(B)$ parameter.

Aiming at obtaining the best fitting to the observed $\mu_B$ in terms
of the globular specific frequencies we then proceeded in two
steps. First, we looked for the $C(B)$ parameter that makes
the quantity $\mu_B+2.5log[N(RGC)+N(BGC)/C(B)]$ constant with
galactocentric radius and, from this constant, after adopting
a given distance modulus and interstellar reddenig, we obtain
$S_B(RGC)$. In turn, $S_B(BGC)$ comes from $C(B)$ and from this
last parameter.

This procedure differs from that given in D2003 in the sense
that:a) Only the profile shape is fitted (colour gradients are
predicted and then compared with observations; b) Constant intrinsic
specific GC frequencies are assumed (i.e., not forcing a particular
functional depedence with galactocentric radius);c) The derived
surface brightness is confronted with an observed profile that
relies on genuine sky brightness measurements (as opposed to the
adoption of a preferred value as in D2003).

In the following sections we describe all the terms involved
in the $\mu_B$ expression that will be used to fit the observed
blue profile.

\section{The NGC 1399 GCS and its distance modulus}
\label{GCSD}

\subsection{Bi-modality}

The possible existence of a bi-modal $(C-T_1)$ colour distribution 
in the NGC 1399 GCS was suggested by \citet{OGF93} and confirmed in 
\citet{OFG98} by means of $(C-T_1)$ vs. $(M-T_1)$ two colour diagrams. This
last work noted that, except in the very inner regions of the
galaxy, the modal colours of both cluster families remained
at constant values independent of galactocentric radius. This
feature was also noticed by D2003 who extended
the survey to about 20 arcmin in radius. As noted in this last
work, and based on \citet{Geisler96}, the $T_1$ and $R_{kc}$ bands
are practically identical and in what follows we adopt
$(C-T_1)$ as equivalent to $(C-R_{kc})$.

\begin{figure}
\includegraphics[width=9cm,height=9cm] {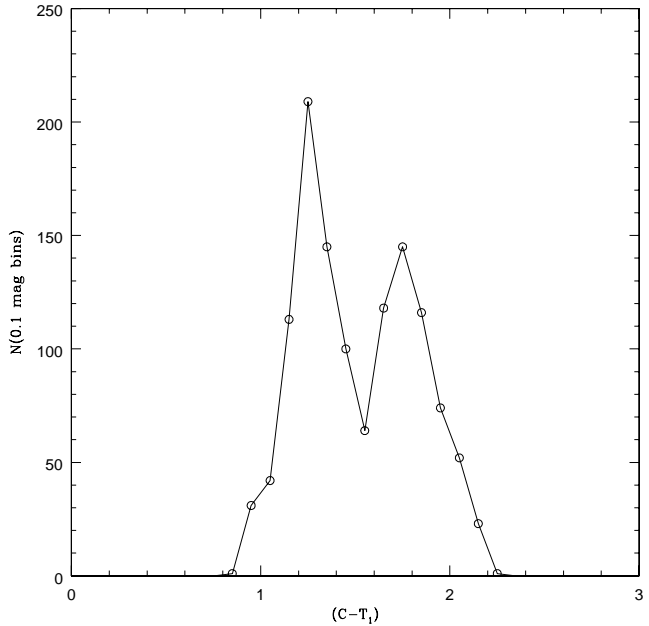}
\caption{$(C-T_1)$ colour distribution for 1246 globular cluster candidates 
associated with NGC 1399 and within a galactocentric radii of 2.0 
to 7.0 arcmin. The sample includes globular candidates with $T_1$ (or $R_{kc}$) magnitudes between 21.0 and 23.2;
colour bins are 0.1 mag wide. Two well defined modal colours are 
seen at $(C-T_1)=1.28$ and 1.77.}
\label{BIMOD} 
\end{figure}

Figure~\ref{BIMOD} shows the $(C-T_1)$ colour distribution for a sample
of 1246 GC candidates contained within an annular region
between 2 and 7 arcmin of the galaxy centre, taken from
the Forte et al. (2002) and D2003 data. The annular boundaries of this region were chosen in order
to maximize the number of GC while keeping a low 
contribution from field interlopers. In turn, the magnitude range
covers from $T_1=21.0$, to the position of the GC
luminosity function turn-over at $T_1=23.2$. We adopt
this value as a compromise between those derived 
by \citet{GF90} and \citet{OFG98}.

GC candidates brighter than $T_1=21$, that were left out of the analysis, 
constitute a small fraction of the whole cluster population 
(less than 9\%) and their nature remains to be clarified as they 
exhibit a uni-modal colour distribution (noted both in \citealt{OFG98},
and D2003) and a higher radial velocity dispersion than that 
characterizing their fainter counterparts (from R2004 data).

A convolution of the original data, that in Figure~\ref{BIMOD} appear
binned in 0.10 mag intervals, with a gaussian kernel with
a dispersion of 0.05 mag (similar to the photometric errors)
leads to $(C-T_1)$ modal colours
of 1.28 and 1.77 for the blue and red globulars, respectively.
In what follows, and consistent with D2003, we adopt a
$(C-T_1)$ colour of 1.55 as a formal boundary between the blue
and red GC (their associated diffuse stellar populations will
also be referenced as ``blue'' and ``red'' in what follows).
The colour excess maps by \citet{SFD98} give
$E(B-V)=0.015$ that translates to $E(C-T_1)=0.030$ along the line
of sight to NGC 1399. The resulting intrinsic colours
are $(C-T_1)_o=1.25$ for the blue GC and
$(C-T_1)_o=1.74$ for the red GC. In order to derive a $(B-R)$ colour
gradient and the specific frequencies of both cluster
families in the $V$ band, $S_n$, by means of the best profile fitting
to the observed $\mu_B$ (Section~\ref{ADP}), we used the following
transformations:

\begin{eqnarray}
(B-V)_o = 0.508(C-T_1)_o + 0.064  \\
(B-R)_o = 0.704(C-T_1)_o + 0.269  
\end{eqnarray}

These last relations were obtained by means of bi-sector fits
to the intrinsic colours of MW globulars \citep*{RHS88}  
that also give a good representation of intrinsic colours in
galaxies with redder globulars (see \citealt{FF2001}) and
yield $(B-V)_o=0.68$, $(B-R)_o=1.13$ for the blue GC and $(B-V)_o=0.95$,
$(B-R)_o=1.49$ for the red GG with an overall uncertainty of
$\approx \pm 0.03$ mag.

\subsection{Distance modulus}

We derive a distance modulus to NGC 1399 by means of the
turn-over magnitude of the blue GC, $B_{to}= 24.32$ mag,
obtained by \citet{GFBE99} on the basis of their
HST photometry. In order to avoid corrections for metallicity
effects (e.g \citealt*{ACZ95}) we compare that
magnitude with the mean blue absolute magnitude of MW globulars
with $(C-T_1)_o$ colours in the same range. The \citet{Harris96} compilation
includes 35 globulars with a mean $(C-T_1)_o=1.15$ (i.e. slightly
bluer than the N1399 mean blue GC colours) and a mean absolute magnitude
$M(B)=-7.15 \pm 0.3$. Adopting $A(B)=0.06$ mag 
(consistent with the $E(B-V)$ excess
mentioned above) leads to $(V-M_{v})_o=31.4$ (corresponding to 19
Mpc). This distance modulus agrees with that adopted by R2004
who give a summary of other distance estimates available in the
literature.

\section{Blue Surface Brightness Profile and $(B-R_{kc})$ colours for NGC 1399}
\label{BandR1399}

The determination of galaxy surface brightness profiles over large
galactocentric radii faces well known problems connected with
the adoption of a proper sky level and flat fielding (particularly
severe in wide field instruments). Small field telescopes, on the
other side, are more manageable in terms of flat field quality
but, in the case of objects with large angular sizes, do not
allow the measurement of a meaningful sky brightness.

In this work we used the 2.15 m telescope at the Complejo
Astron\'omico El Leoncito (San Juan, Argentina) in combination
with a focal reducer that gives an effective field of 9
arcmin on the CCD detector. Simultaneous and continuous measurements
of the sky brightness were obtained by means of a zenithal airglow
sensor \citep{Scheer87} that measures the 
intensity of atmospheric molecular emission lines. These measurements
were correlated with $B$ and $R_{kc}$ sky brigtness estimates made
on CCD frames taken 40 arcmin to the north of the galaxy centre (and intertwined
with the galaxy field images).

The correlation
took into account the differences in airmass between the zenith
and the line of sight to the galaxy as well as atmospheric
extinction (derived during the run: $k_B=0.29$; $k_R=0.08$).
The sky brightness to be subtracted from each galaxy field was
then obtained by interpolating its temporal beahviour derived from
the airglow vs. sky frames correlation.

Sky flats were obtained inmediately after sunset and provided
an overall flat field on the order of $0.5 \%$.

The observation sequence included sets of four images (15 min exposures)
centreed on the galaxy and off set frames taken along a position angle
of 45 degrees (N to E) and reaching a galactocentric distance of
about 12 arcmin. This direction is clean of bright galaxies that
may contaminate the NGC 1399 outer profile.

Typically, the $B$ sky magnitude ranged from 23.1 to
23.3 mag per square arcsec during the night. We note that the
observing run took place in December 1997 and during a period of
exceptionally low airglow level coincident with a minimum of solar
activity.

Images centered on the galaxy were processed with the ELLIPSE routine
within IRAF\footnote{IRAF is distributed by National Optical Astronomy Observatories, which are operated by the Association of Universities for Research in Astronomy, Inc., under co-operative agreement with the National Science Foundation.} and delivered brightness profiles that reach 240
arcsec from the galaxy centre. The blue profile was extended
up to 510 arcsec using three 20 minute exposure images obtained with the 4m Blanco 
Telescope at CTIO (described in \citealt{Forte2002}). These last images are not suitable for
an estimate of the sky level and where then matched to our own profiles by means of the method described by \citet{CCD94}.

The much higher sky brightness in the $R_{kc}$ band prevented a similar
approach, restricting the limiting galactocentric radius
to 240 arcsec with a maximum estimated error of $\pm 0.04$ mag at that
radius.
       
In the case of the off-centre blue frames, aperture
photometry was carried out within 25 square windows (5 arcsec
on a side) located on a region of the images were the flat field 
was considered better than 0.2\% and far from bright stars 
(in order to avoid spurious scattered light). These measures
provide the surface brightness at mean galactocentric radii of 410, 470,
(overlapped with the extended profile), 605 and 730 arcsec.

\begin{figure}
\includegraphics[width=9cm,height=9cm] {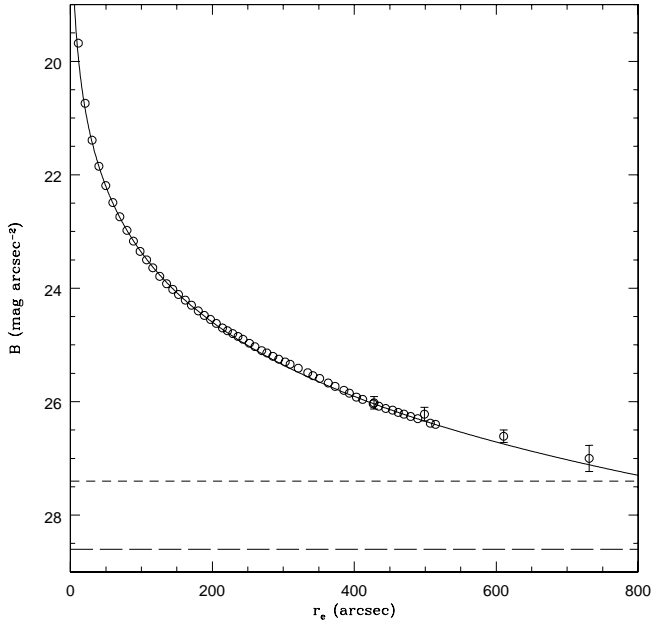}
\caption{Blue surface brightness profile for NGC 1399 as a function of
elliptical radius ($r_e=\sqrt{ab}$). The bars correspond to the error
of the mean magnitude. Errors for galactocentric radii smaller than
400 arcsec are comparable to the size of the plotted symbols ($\pm 0.03$ mag.)
The solid line is a S\'ersic profile. The
long dashed and short dashed lines correspond to one $\sigma$ and
to three $\sigma$ uncertanties of the sky level.}
\label{BPROF} 
\end{figure}

The final composite blue profile is shown in Figure~\ref{BPROF} and listed
in Table 1 where $(B-R_{kc})$ colour indices are also given up to
an elliptical  galactocentric radius of 228.8 arcsec.
The elliptical galactocentric radius ($r_e=\sqrt{ab}$) was computed
using the ellipticities delivered by ELLIPSE and adopting a
constant ellipticity $\epsilon =(1-b/a)=0.13$ for the two outermost
off-set frames. Error bars depict the error of the mean magnitude derived
on the outer CCD frames.

\begin{table}
 \caption{NGC 1399 Surface Brightness Photometry}
\begin{tabular}{|r|c|c|l|l|}
\hline
\multicolumn{1}{c}{r} &
\multicolumn{1}{c}{$log r$} &
\multicolumn{1} {c} {$r^{1/4}$}  &
\multicolumn{1} {l} {$~B$}    &
\multicolumn{1} {l} {$B-R_{kc}$} \\
\multicolumn{1}{c}{arcsec} &
\multicolumn{1}{c}{ } &
\multicolumn{1}{c}{ }  &
\multicolumn{1} {c} { }  &
\multicolumn{1} {c} { } \\
\hline
    11.20  &   1.049  &    1.829 &   19.68   & 1.58 \\
    20.80  &   1.318  &   2.136  &   20.74   & 1.53 \\
    30.50  &   1.484  &   2.350  &   21.39  &  1.51 \\
    40.10  &   1.603  &   2.516  &   21.85  &  1.50 \\
    49.80  &   1.697  &   2.656  &   22.19  &  1.49 \\
    59.80  &   1.777  &   2.781  &   22.49  &  1.51 \\
    69.60  &   1.843  &   2.888  &   22.74  & 1.49 \\
    79.70  &   1.901  &   2.988  &   22.98  &  1.48 \\
    88.80  &   1.948  &   3.070  &   23.17  &  1.48 \\
    98.20  &   1.992  &   3.148  &   23.35  &  1.46 \\
   107.30  &   2.031  &   3.218  &   23.50  &  1.47 \\
   116.10  &   2.065  &   3.283  &   23.64  &  1.47 \\
   125.80  &   2.100  &   3.349  &   23.79  &  1.46 \\
   135.50  &   2.132  &   3.412  &   23.92  &  1.47 \\
   144.20  &   2.159  &   3.465  &   24.02  &  1.45 \\
   152.30  &   2.183  &   3.513  &   24.11  &  1.45 \\
   161.90  &   2.209  &   3.567  &   24.21  &  1.43 \\
   170.60  &   2.232  &   3.614  &   24.30  &  1.44 \\
   180.20  &   2.256  &   3.664  &   24.40  &  1.44 \\
   188.70  &   2.276  &   3.706  &   24.48  &  1.41 \\
   197.40  &   2.295  &   3.748  &   24.55  &  1.41 \\
   205.60  &   2.313  &   3.787  &   24.62  &  1.41 \\
   214.30  &   2.331  &   3.826  &   24.70  &  1.40 \\
   221.40  &   2.345  &   3.857  &   24.75  &  1.40 \\
   228.80  &   2.359  &   3.889  &   24.80  &  1.41 \\
   236.20  &   2.373  &   3.920  &   24.85  &      \\
   243.20  &   2.386  &   3.949  &   24.90  &      \\
   252.20  &   2.402  &   3.985  &   24.97  &      \\
   260.00  &   2.415  &   4.016  &   25.03  &      \\
   269.60  &   2.431  &   4.052  &   25.10  &      \\
   276.80  &   2.442  &   4.079  &   25.14  &      \\
   285.30  &   2.455  &   4.110  &   25.20  &      \\
   293.50  &   2.468  &   4.139  &   25.25  &      \\
   302.70  &   2.481  &   4.171  &   25.30  &      \\
   309.60  &   2.491  &   4.195  &   25.34  &      \\
   321.10  &   2.507  &   4.233  &   25.41  &      \\
   334.40  &   2.524  &   4.276  &   25.49  &      \\
   341.70  &   2.534  &   4.299  &   25.54  &      \\
   351.30  &   2.546  &   4.329  &   25.59  &      \\
   363.30  &   2.560  &   4.366  &   25.67  &      \\
   373.00  &   2.572  &   4.395  &   25.73  &      \\
   385.40  &   2.586  &   4.431  &   25.80  &      \\
   393.10  &   2.595  &   4.453  &   25.85  &      \\
   403.00  &   2.605  &   4.480  &   25.92  &      \\
   411.80  &   2.615  &   4.505  &   25.96  &      \\
   426.80  &   2.630  &   4.545  &   26.04  &      \\
   427.80  &   2.631  &   4.548  &   26.02($\pm 0.11$)  &      \\
   434.50  &   2.638  &   4.566  &   26.08  &      \\
   444.30  &   2.648  &   4.591  &   26.12  &      \\
   454.00  &   2.657  &   4.616  &   26.15  &      \\
   462.00  &   2.665  &   4.636  &   26.19  &      \\
   470.00  &   2.672  &   4.656  &   26.22  &      \\
   479.40  &   2.681  &   4.679  &   26.26  &      \\
   489.50  &   2.690  &   4.704  &   26.30  &      \\
   498.90  &   2.698  &   4.726  &   26.22($\pm 0.12$)  &      \\
   507.10  &   2.705  &   4.745  &   26.38  &      \\
   514.60  &   2.711  &   4.763  &   26.40  &      \\
   610.50  &   2.786  &   4.971  &   26.61($\pm 0.11$)   &      \\
   731.00  &   2.864  &   5.200  &   27.00($\pm 0.23$)   &      \\
\hline
\end{tabular}
\end{table}


 Figure~\ref{BPROF} also includes, as a reference, the fit
of a S\'ersic law,  $\mu_B=b_o+b_1(r_e/\alpha)^n$, 
with $b_o=3.74$, $b_1=1.086$, $\alpha=4.82 \times 10^{-13}$ and $n=0.0878$.

\begin{figure}
\includegraphics[width=9cm,height=9cm] {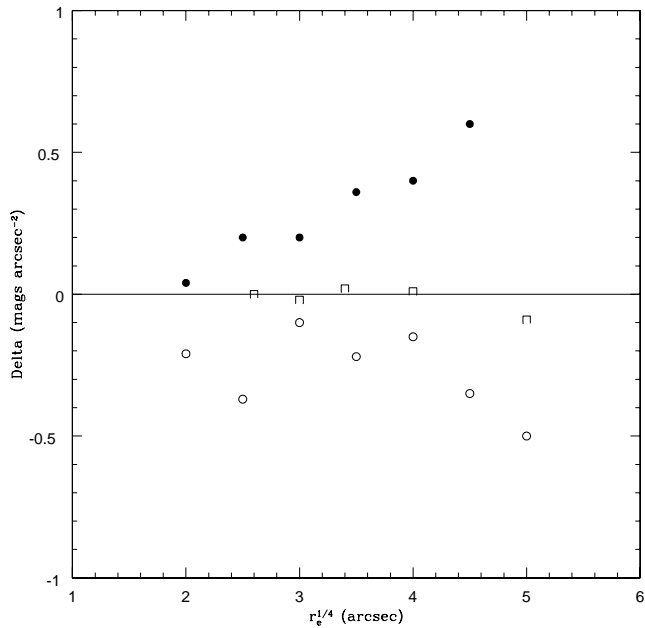}
\caption{Comparison between the blue profile presented in this work with
those given by \citet{KB88} (open circles), \citet{CCD94} (filled circles) 
and \citet{HIR99} (squares). This last authors $V$ profile was 
transformed adopting a constant $(B-V)=0.93$.}
\label{RESID}
\end{figure}

Large field blue surface brightness profiles for NGC 1399 have been 
published by \citet{KB88} (who combine their observations with
$V$ values from \citealt{Scho86}) and by \citet{CCD94}. A
comparison between these profiles shows marked differences that
increase with galactocentric radius. Figure~\ref{RESID} shows the brightness residuals as a
function of ${r_e}^{1/4}$ when comparing the profiles of those authors
with ours (that, approximately, falls in the middle of both).
This diagram also includes a comparison with the $V$ brightness
profile presented by \citet{HIR99}, by adopting a constant
difference $(B-V)=0.93$, indicating a good overal agreement (although
some residual differences might be expected if a colour gradient
were present).

D2003 present $R_{kc}$ and $C$ profiles that reach some 20 arcmin in
      galactocentric radius. These profiles were obtained by adopting
      a preferred sky brightness on the basis of measurements
      at the edge of the images. In particular, their $R_{kc}$ profile is
      referred to again in Section~\ref{RES} .

\section{The Areal Density Profiles}
\label{ADP}
    Numerous studies of GCS in luminous elliptical galaxies agree
    in the sense that the behavior of the areal densities of the the blue
    and red GC show a similar behaviour from galaxy to galaxy: blue clusters
    usually show a more shallow distribution than that corresponding to the
    red ones which, in turn, show density slopes more similar (but not
    always identical) to that of the galaxy haloes. A custommary approach
    uses $r^{1/4}$ laws to fit the density run with galactocentric radius
    although these approximations fail in the inner regions where GCS
    exhibit flat cores (see for example, \citealt{KW98}; \citealt*{LLG2002}). These core profiles show a marked difference with the galaxy
    light profiles. This last feature has been used to estimate the rate of GC
    destruction due to gravitational effects in NGC 1399 and in other
    galaxies (\citealt{CDD2001};\citealt{CDT99}).

\begin{figure}
\includegraphics[width=9cm,height=9cm] {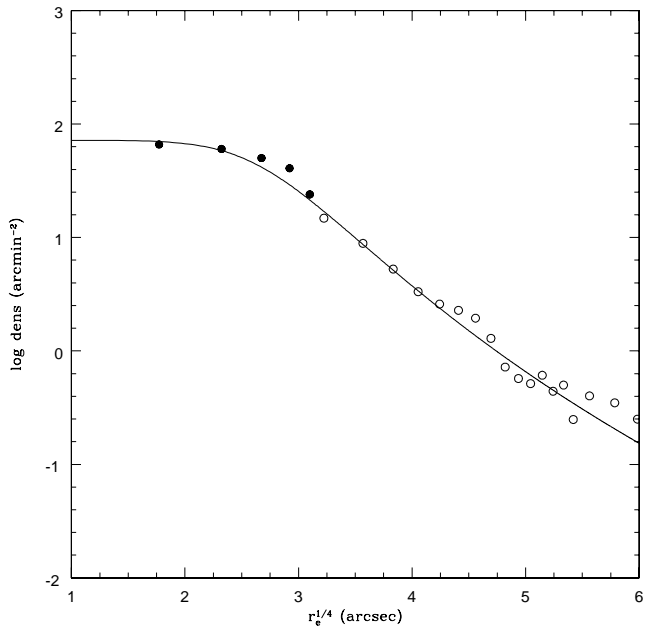}
\caption{Areal density of the NGC 1399 red globular clusters. Open circles from
       \citet{D2003} data. Filled circles from the (scaled) density
       profile given by \citet{FBG98}. The solid line is the
       approximate fit of a Hubble law with $r_c=60$ arcsec and a central
       density of 72 globulars per sq. arcmin (down to magnitude R=23.2).}
\label{RGCP} 
\end{figure}

    Figure~\ref{RGCP} depicts the areal density of the red GC derived from D2003 data.
    An attempt to derive the ellipticity of the red clusters distribution 
    (and also for the blue ones) based on azimuthal counts,
    indicates a value close to zero (comparable with that of the galaxy
    itself), which we adopted  in what follows.
    The inner values (filled circles) were derived from HST observations
    given by \citet{FBG98}. Their counts were normalized to
    a limiting magnitude $T_1=23.2$ and scaled down by $2/3$. This last factor
    takes into account that only this fraction of the GC seem to
    belong to the red population on the basis of the colour statistics
    in the inner region of the galaxy, also based on HST observations,
    presented by \citet{GFBE99}.

    For comparison purposes only, Figure~\ref{RGCP} also shows a Hubble law 
    characterized by a central density of 72 GC per sq. arcmin and a core 
    radius of 60 arcsec.

\begin{figure}
\includegraphics[width=9cm,height=9cm] {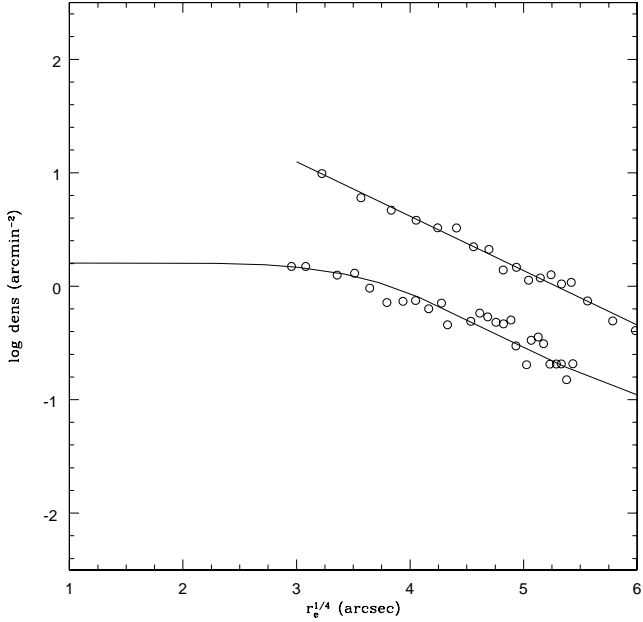}
\caption{Areal density of the NGC 1399 blue globular clusters. The straight 
       line is the best formal fitting with an ${r}^{1/4}$ law.  The lower set
       of values corresponds to the bluest globulars (see text) fit with an 
       isothermal profile with a core radius of 275 arcsec.}
\label{BGCP} 
\end{figure}

    The blue GC density profile is depicted in Figure~\ref{BGCP} 
    that also includes the best formal fitting with an ${r}^{1/4}$ law.
    In this case we note that there is a marginal trend in the sense that
    the bluest clusters($(C-T_1)=0.90$ to $1.20$, also shown in the same figure) 
    exhibit a shallower core. As a reference we tentatively fit an isothermal
    profile \citep{BT87} with a core radius of 275 arcsec
    for these objects. However this feature will require further confirmation
    using a deeper sampling and, in what follows, we consider the blue GC
    as a single family.

    One important aspect related with the determination of the GC
    density profiles at large galactocenric radii is the impact of unresolved
    field interlopers with colours similar to those of the clusters. We
    stress that D2003 data were processed with a highly
    successful procedure for the identification of non-resolved sources,
    as the GC are expected to be at the Fornax cluster distance. Further
    spectroscopic data presented in R2004 indicate a contamination level
    on the order of 10 \% or less. In addition,
    their comparison field (30 arcmin on a side) located
    three degrees towards the NE of NGC 1399, provides a very solid
    estimate of the background level in terms of the uncertainties of the
    statistical counts (see Section 7).

\section{The GC Volumetric Density Profiles}
\label{VDP}
    Volumetric density profiles allow an estimate of the total mass within
    a given spatial galactocentric radius as well as a comparison with
    other galaxy components (e.g. hot gas, dark matter) and, also, to
    estimate the effect of an eventual (spatial) colour gradient 
    on the projected integrated colours. We note that neither the
    galaxy light profile nor the GC areal densities can be approximated
    by means of single (spatial) power laws ($r^\alpha$) except for limited
    ranges in galactocentric radius. In this work, and following R2004, we adopt
    integrable volumetric density functions of the form:

\begin{equation}
Dens(r)= \rho_{o} \left({\frac{r_s}{r}}\right) ^\zeta \left( {\frac{1}{1+\frac{r}{r_s}}}\right) ^{3-\zeta}
\end{equation}
\noindent
    and aim to determine the $\zeta$ and $r_s$ parameters that, once Dens(r)
    is integrated along the line of sight, give a good representation of the
    projected (or areal) density profiles. The integration was carried out
    within a limiting radius equal to the virial radius derived by
    R2004, $R_{vir} \approx 5800$ arcsec. We note, however, that
    changing this limit by a considerable amount does not have a
    large effect on the results of the integration.

\begin{figure}
\includegraphics[width=9cm,height=9cm] {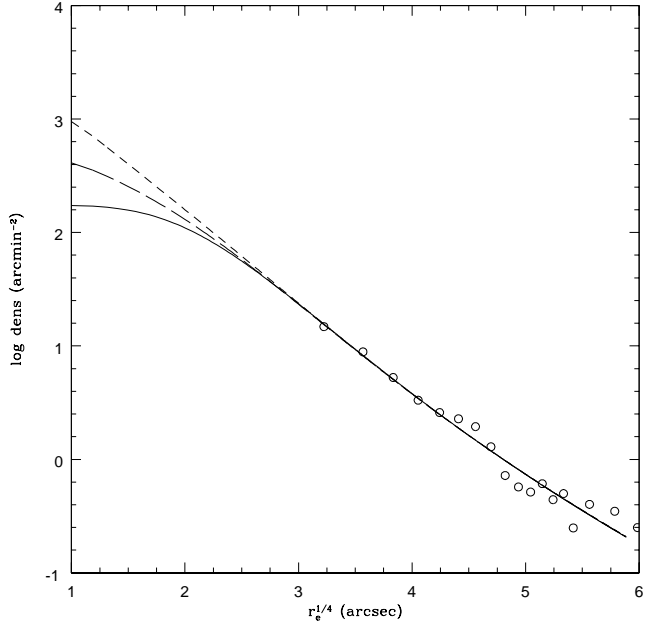}
\caption{Projected volumetric density profiles that give an adequate fit
       to the observed red GC surface density (circles) for a given 
       combination of $r_s$ and $\zeta$ values (see text); $\zeta=1.5$, $r_s=70$ arcsec
       (short dashed); $\zeta=1.0$, $r_s=50$ arcsec (long dashed); $\zeta=0.0$, $r_s=30$ arcsec
       (solid line). The profiles overlap for galactocentric radii larger than
       80 arcsec.}
\label{RGCFIT} 
\end{figure}

    The best formal fits to the areal density of the red GC are displayed
    in Figure~\ref{RGCFIT}. The results show that
    different combinations of the $\zeta-r_s$ parameters can yield almost equivalent fits
    within the range of the fit (120 to 1200 arcsec) and that
    this degeneracy breaks up only in the innermost regions of the
    galaxy. However, deciding which of these profiles might be the best
    representation of the \textit{original} GC distribution is probably
    impossible as the disrupting gravitational
    effects are expected to be important in these regions.

\begin{figure}
\includegraphics[width=9cm,height=9cm] {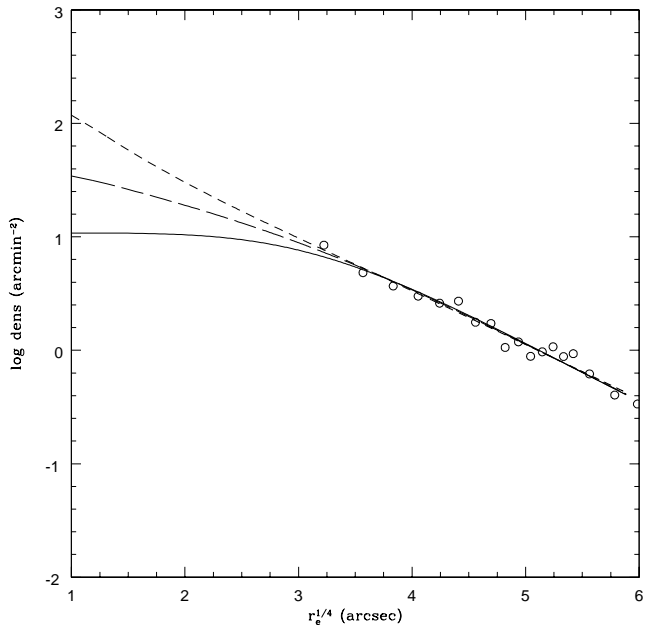}
\caption{Same as Figure 6 but for the blue GC (open circles). $\zeta=0.0,
        r_s=200 $ arcsec (continuous line); $\zeta=1.0$, $r_s=400$ arcsec (long
        dashed); $\zeta=1.5$, $r_s=700$ arcsec (short dashed). Profiles
         overlap for galactocentric radii larger than 150 arcsec.}
\label{BGCFIT} 
\end{figure}

    Figure~\ref{BGCFIT} shows the same kind of diagram for the blue GC. 

\section{Results}
\label{RES}
     This section presents the results of fitting the composite
     stellar population profile described in Section~\ref{GCSD} to the
     blue surface brightness presented in Section~\ref{BandR1399}. The quality
     of the fit is estimated in terms of:
\begin{itemize}

\item  \textit{Shape of the surface brightness profile}

\item  \textit{Colour gradient}

\item  \textit{Cumulative GC specific frequency}

\end{itemize}

\subsection{The shape of the surface brightness profile}

     In order to derive the best fitting to the observed blue surface
     brightness profile (defined as the one that minimizes the square
     of the brightness residuals) we proceeded as folows:

\begin{itemize}

\item [-] A grid of volumetric density profiles that, once projected, provide 
     acceptable fits (with residuals within 1.5 times the mimimum dispersion) to 
     the observed areal densities was generated for each
     GC family (setting $\zeta=$0.0, 1.0 and 1.5) and changing the $r_s$
     parameter.

\item [-] These profiles were projected on the sky (from r=0 to 1200
     arcsec) and then combined within the $\mu_B$ equation (described
     in Section~\ref{CSBP}). The constant $C(B)$ was iteratively changed until
     the quantity $k=\mu_B+2.5log [N(RGC)+N(BGC)/C(B)]$ remained constant
     with galactocentric radius, yielding a value of 33.70. The mimimum 
     dispersion of this quantity ($\pm 0.035$) was obtained by combining GC density 
     profiles characterized by $\zeta=1.5$ and $r_s=30$ arcsec for the red 
     GCS, and $\zeta=1.0$ and $r_s=375$ arcsec for the blue GC.

\item [-] The $k$ value was equated to $(V-Mv)_o+A(B)+2.5log[S_B(RGC)]$
     and, by adopting the corrected distance modulus and interstellar
     extinction discussed in Section~\ref{GCSD}, the instrinsic specific
     frequency of the red GC was derived. The specific frequency of the
     blue clusters follows from $C(B)$ and $S_B(RGC)$.
\end{itemize}

     This procedure leads to $C(B)=3.5 \pm 0.5$, $S_B(RGC)=7.8$, $S_B(BGC)=27.3$.
     The uncertainty for the $C(B)$ parameter comes, mainly, from the
     range of values allowed by the errors in the outermost region of the
     brightness profile.
     The resulting brightness profile is compared with the observations
     in Figure 8 and the fit residuals are shown in Figure 9. This last
     diagram indicates that, from 30 to 720 arcsec the approach gives
     a very good fit to the observations with maximum deviations
     of about $\pm 0.05$ mag and an overal rms of $\pm 0.035$ mag.

     The inner 30 arcsec show a deviation in the sense that the
     nuclear region is brighter than the model. Improving the fit
     in this region might require a higher $\zeta$ exponent (and
     also the consideration of seeing effects). However, as discussed later,
     some arguments support the presence of a distinct stellar population
     in the galaxy nucleus.

     The volumetric profiles assumed a spherical geometry. Even though
     the nature of the residuals in Figure 9 cannot be completely
     asserted, we note that they seem correlated with the ellipticity
     variations depicted in the upper curve of that figure. The assumption
     of an oblate geometry did not improve the fit suggesting the
     necessity of a three axis ellipsoid for a further comparison.

     We stress that, due to the
     small fractional contribution of the blue stellar population to the
     integrated light in the inner regions of the galaxy, the fit is rather
     insensitive to the set of $r_s-\zeta$ values adopted for the blue GC. The
     reason for adopting the quoted values are commented on in Section 11.

\subsection{Sources of uncertainty}

        One source of uncertainty comes from the background level
        adopted as representative of the field contamination by non-resolved
        objects with colours in the domain of the GC colours. The
        comparison field used by D2003 indicates values of 0.31 and 0.08
        objects per sq. arcmin for the blue and red GC respectively.
        This field (approx. 35 by 35 arcmin) includes
        a large number of objects and yields low counting uncertainties
        ($\pm 0.016$ and $\pm 0.008$). However, the doubt remains in the sense if
        this is in fact representative of that of the central regions of
        the Fornax cluster (three degrees away to the SW).

        An increase (decrease) of the background levels by $50 \%$
        of the adopted values changes the shape of the density (and
        brightness) profiles forcing an increase (decrease) of the
        $C(B)$ ratio of $\approx 0.5$ in order to keep an acceptable
        fit to the observed profile.

        The combined uncertainty of the quantities involved ($k:\pm 0.035$);
        $A(B):\pm 0.02$; $(V_{o}-M_{v}):\pm 0.1$) lead to $\pm 0.8$ in
        the $S_{B}(RGC)$ parameter. This value, and the quoted uncertainty
        on the $C(B)$ ratio, in turn yield $\pm 4.8$ for $S_{B}(BGC)$.

\begin{figure}
\includegraphics[width=9cm,height=9cm] {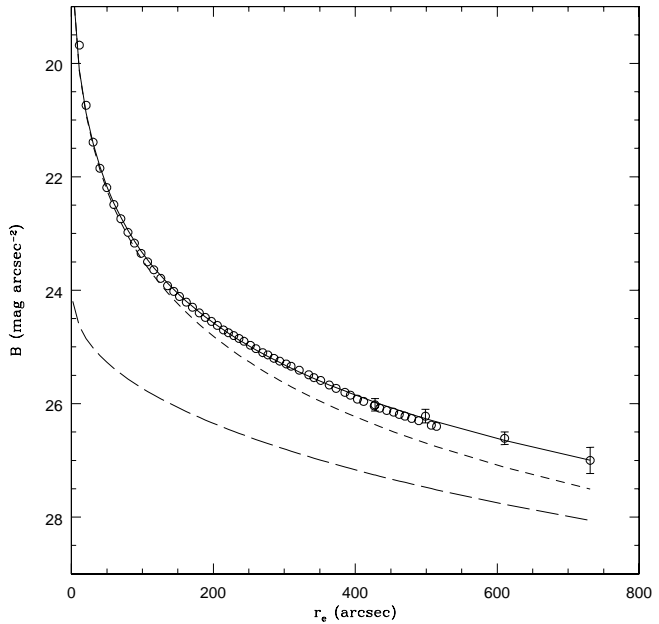}
\caption{Best profile fit to the observed blue surface brightness of NGC 1399
       (solid line) The short dashed line corresponds to the light 
       contribution of the ``red'' stellar population; the long dashed line
       is the light contribution from the ``blue'' stellar population.
       The associated GC have $S_n(RGC)=3.3$ and $S_n(BGC)=14.3$}
\label{BESTFIT} 
\end{figure}

\begin{figure}
\includegraphics[width=9cm,height=9cm] {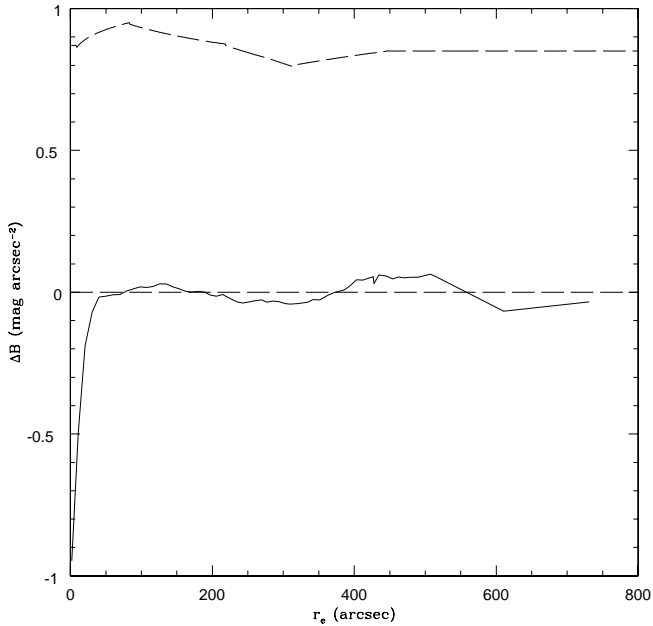}
\caption{Blue profile fit residuals as a function of the elliptical 
       galactocentric radius $r_e$ (solid line). The maximum deviations
       are within $\pm 0.05$ mag with an overall rms of $\pm 0.035$ mag over
       a galactocentric radius from 30 to 730 arcsec. Note the marked
       deviation within 30 arcsec. The upper (dashed) curve shows
       the variations of the galaxy flattening ($q=b/a$). A constant q
       value (0.83) was adopted for $r_e$ larger than 420 arcsec.}
\label{BESTRES} 
\end{figure}

\subsection{$(B-R_{kc})$ colour gradient}

\begin{figure}
\includegraphics[width=9cm,height=9cm] {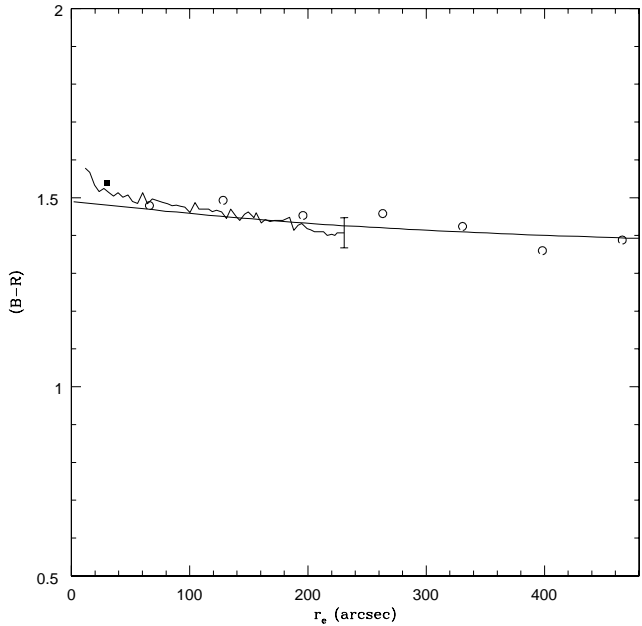}
\caption{Galaxy $(B-R_{kc})$ colour as a function of elliptical galactocentric
       radius $r_e$ (solid curve). The straight line is the expected gradient
       from the model described in Section 2. The square indicates the
       $(B-R_{kc})$ colour measured by \citet{Mich2000} at $r_0=22.8$ arcsec.
       The bar at 228.8 arcsec indicates the uncertainty of our colour
       colour gradient. Open circles show the resulting colour gradient
       obtained by combining our photometry and that given in D2003
       (see text).}
\label{colour} 
\end{figure}

     A composite red profile was computed by adding the red light contribution
     of each stellar population (adopting the
     $(B-R_{kc})$ colour indices and colour excess discussed in Section~\ref{GCSD}). 
     The resulting $(B-R_{kc})$
     colour gradient is displayed in Figure~\ref{colour}, that
     also includes the observed values listed in Table 1. This diagram
     shows that, at galactocentric radii larger than 30 arcsec,
     the predicted colours are consistent with the observations.
     In turn, these observations are in very good agreement with
     \citet{Mich2000}. This last author gives, at a galactocentric radius
     $r_0=22.8$ arcsec, $(B-R_{kc})=1.54$ (0.02 mag redder than ours) and a colour
     gradient $dlog(B-R_{kc})/dlog(r)=-0.05$ that compares well both with the
     observed ($-0.06 \pm 0.02$) and with the predicted (-0.045) colour
     gradient.

     Another estimate of the $(B-R_{kc})$ gradient was obtained by combining
     our blue profile and the $R_{kc}$ one given in D2003. These last
     authors adopted a ``preferred'' sky level and warned about the
     high sensitivity of the profile at the outermost radius. The uncertainty 
     of their profile 
     becomes $\approx \pm 0.05$ mag within a galactocentric radius 
     of 8 arcmin that we adopt as the boundary of our combined $(B-R_{kc})$ colour
     profile. The resulting $(B-R_{kc})$ is also depicted in Figure~\ref{colour} after
     applying a zero point shift of -0.1 mag. This correction is
     necessary in order to bring the resulting colours into agreement with
     \citet{Mich2000} and with ours.

     As a summary, both our observed $(B-R_{kc})$ gradient and that derived 
     by combining our blue profile with the red one from D2003, are
     consistent with that predicted by the profile fit.

\subsection{The intrinsic and cumulative GC specific frequencies}

\begin{figure}
\includegraphics[width=9cm,height=9cm] {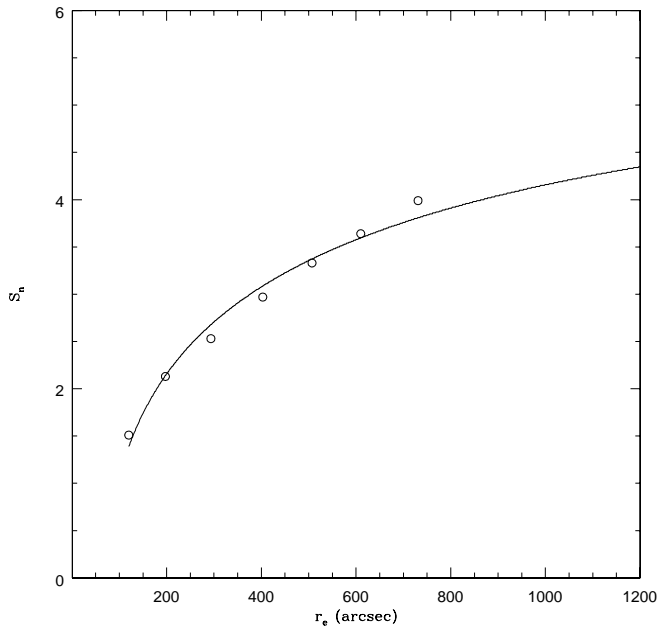}
\caption{Cumulative specific frequency of the GCS associated
       with NGC 1399. Open circles represent $S_n$ values derived from the
       actual counts (transformed to total GC population; see text)
       combined with the $V$ surface brightness profile predicted by the
       model. The solid line comes from both the fitted GC areal 
       densities and the model $V$ profile.}
\label{Sn} 
\end{figure}

     The intrinsic specific frequencies derived for each GC family were transformed
     to the $V$ band, adopting the $(B-V)_o$ colours given in Section~\ref{GCSD}, yielding
     $S_n(RGC)=3.3 \pm 0.3$ and $S_n(BGC)=14.3 \pm 2.5$. In the case of the
     red GC, the derived specific frequency is coincident with the representative
     value quoted for normal ellipticals (see \citealt*{HHM98}).
     The composite cumulative (with galactocentric radius) specific frequency
     was computed adopting a total number of 700 GC within a galactocentric
     radius of 120 arcsec (from the density curve given by \citealt{FBG98}). GC counts at larger radii were derived fom D2003
     data adopting a limiting magnitude $T_1=23.2$, a fully gaussian
     integrated luminosity function with a turn-over at this magnitude,
     and an overall completeness of 90 \%.
     In turn, the galaxy $V$ surface brightness profile was derived from
     our observed blue profile and the $(B-V)$ colour gradient computed from
     the model fit.
     The adoption of the distance modulus and interstellar extinction
     discussed in Section~\ref{GCSD} lead to the $S_n$ values shown in Figure~\ref{Sn}.
     The derived cumulative $S_n$ is somewhat smaller, but comparable
     within the uncertainties, with results given in D2003 and in
     previous works (\citealt{OFG98}; \citealt{Forte2002}) that
     have pointed out that the NGC 1399 GCS cannot be considered as
     an anomalously high $S_n$ system.

\section{A Caveat about the approach}
\label{UNCER}
        
        An alternative fitting to the observed surface 
        brightness profile can be obtained by assuming that the stellar halo 
        follows a density profile similar to that of the red GC and 
        \textit{forcing} a spatial colour gradient $d(B-R_{kc})/d(log r)=-0.045$. 
        
        This blueward trend with increasing galactocentric radius, if due to 
        a change of metallicity, would increase the $B$ brightness of the
        stellar population (at constant mass) in the outer regions. The
        luminosity increase was estimated from \citet{Wor94} population
        synthesis models and assuming an age of 12 to 17 Gy (the results are not 
        very sensitive to age in this regime). The forced spatial gradient 
        also provides a good fit to the observed projected $(B-R_{kc})$ colour 
        gradient. 

        If that gradient were also shared by the red GC, their (projected
        on the sky) number vs. colour statistic, computed on the basis of the
        volumetric density profile, would show a modal colour shift of -0.14 mag
        in $(B-R_{kc})$, and close to -0.20 mag in $(C-T_1)$, over a galactocentric
        radius of 20 arcmin. This result is in conflict with the observations
        (e.g. D2003) that show a practically constant modal colour with
        galactocentric radius for both GC populations.
        
        Despite this last comment, the main caveat about the profile 
        fit approach presented in this work is the possibility that the 
        chemical enrichment history of the red field stellar population 
        might be in fact decoupled from that of the red GC.
        In fact there are theoretical reasons to expect some degree of
        decoupling. For example, if GC form at the beguining of the starburst
        when external pressure and content of gas is highest; as the star
        formation rate (SFR) decreases, chemical enrichment may continue
        to ocur (but no new GC are formed). This would presumably lead to
        a color difference between GC and the galaxy halo. We note, however,
        that in the case of the halo, the result will be luminosity weighted,
        i.e., the effect on the color will depend upon the ratio of the
        number of stars formed during the SFR peak and those formed
        subsequently. If this ratio is high, the resulting halo color
        could be only slightly redder than those of the RGC.

\section{The Stellar Metallicity Distribution Function}
\label{SMDF}

    This section presents a tentative picture about the expected stellar
    metallicity distribution function (MDF) based on the two component
    model presented above.
    Observations of resolved stars in the halo of NGC 5128     
    show marked differences of their metallicity distribution when
    compared with that of GC in number by number statistics
    \citep{HH2002}.

    If in fact the GC specific frequency is the result of a composite
    population, as suggested by the previous section,
    an estimate of the number ratio of metal rich (red population) to
    metal poor (blue population) stars can be derived through the
    intrinsic specific frequencies and their mass to light    ratios.

    The expected stellar mass associated with $N$ clusters within a
    (projected or spatial) galactocentric radius is $M(stellar)=(M/L)(N/S_n)$.
    For the red population we adopt a blue $(M/L)_{red~pop}=10$ derived by \citet{Sag2000} for the inner regions of NGC 1399 (where that population
    is dominant) on the basis of their kinematics study.
    By assuming that the colour difference between both stellar populations
    arises, mostly, from a metallicity difference close to 1.0 dex (see below),
    and the results from stellar population modelling (e.g. \citealt{Wor94}),
    we get $(M/L)_{blue~pop}=5.75$ for the blue population. This ratio is consitent with
    the fact that the turn over of the integrated luminosity function of
    the blue GC in the blue band is about 0.4 mag brighter than that
    of the red GC as determined by \citet{GFBE99}.

    The resulting stellar mass ratio then is:
\begin{eqnarray}
\frac{M(Red)}{M(Blue)}= \frac{S_B(BGC)}{S_B(RGC)} \frac{(M/L)_{red~pop}}{(M/L)_{blue~pop}}\frac{N(RGC)}{N(BGC)}
\end{eqnarray}
     In the particular case of the region between 120 and 420 arcsec
     of galactocentric radius (see Figure~\ref{BIMOD}) where 
     $N(RGC) \approx N(BGC)$,
     the adoption of the specific frequencies derived in Section~\ref{RES},
     leads to a mass ratio close to 6. If the stellar mass distribution
     function of both populations do not differ markedly, this result
     implies that blue stars will be relatively inconspicuous in a number
     statistic that will be largely dominated by the red population stars.

     The shape of the MDF for these stars can the be estimated by transforming the GC
     colours to metallicity and combining both populations according to
     the mass scaling factor derived above. This procedure deserves a cautious
     comment taking into account several factors that may have
     an impact on colour-metallicity relations, like the morphology
     of the horizontal branch stars \citep{GLK96} or the existence
     of an age spectrum among the NGC 1399 GC (Forbes et al. 2001).

     Integrated $(C-T_1)$ colours for the GC sample shown in Figure~\ref{BIMOD}, 
     that covers a galactocentric radius from 11 to 38 Kpc, were
     transformed to metallicity by means of the calibration presented
     by \citet{HH2002}

\begin{eqnarray}
[Fe/H]=-6.037 \big[ 1.0 - 0.82(C-T_1)_o + 0.162{(C-T_1)_o}^2 \big]
\end{eqnarray}

\noindent that improves the linear relation given by \citet{GF90}.
     The GC sample was divided in two groups: blue GC with $(C-T_1)=0.9$
     to 1.60 and red GC with $(C-T_1)=1.5$ to 2.2 (i.e. allowing a 0.1
     mag overlap between both samples) and correcting for interstellar
     reddening.

\begin{figure}
\includegraphics[width=9cm,height=9cm] {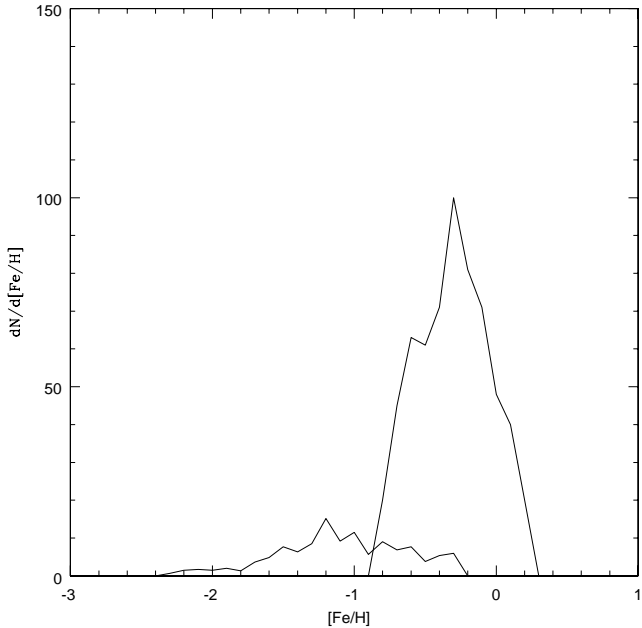}
\caption{Expected metallicity distribution function (MDF) for the stellar
       population within a (projected) galactocentric radius of
       2 to 7 arcmin whose colour distribution is shown in Figure~\ref{BIMOD}.
       The predicted composite MDF shows a good qualitative agreement with
       observations of resolved halo stars in NGC 5128 \citep{HH2002},                 
       i.e. a dominant and wide high metallicity population and
       a less conspicuous and extended blue tail.}
\label{MDFPLOT} 
\end{figure}

     The resulting MDF (in relative units), after applying the scaling factor 
     described before, is shown in Figure~\ref{MDFPLOT}. This distribution 
     shows a 
     good qualitative agreement with that corresponding to the outer
     fields in NGC 5128 from \citet{HH2002}, i.e, a dominant red 
     population with
     a tail extending to low metallicity values.
     The results of these last authors stand on color magnitude diagrams
     of resolved stars interpreted in terms of evolutionary tracks with
     different ages and metallicities. In turn, the MDF presented in
     this work relies on an empirical integrated-colors vs. metallicity
     relation. As a consequence, any further quantitative comparison will
     require a previous effort in order to check for the consistency
     of both approachs in terms, for example, of the metallicity scale.

\section{Stellar mass and GC formation efficiency}
\label{SMEFF}

    The GC formation efficiency is a key parameter in the understanding
    of the link between these clusters an the diffuse stellar population
    of the halo. For a given GC specific frequency, the formation
    efficiency, in terms  of the associated stellar mass, is  

\begin{eqnarray}
\eta=\frac{S_{n}(\lambda)M_{gc}}{(M/L)c}
\end{eqnarray}

\noindent where (M/L) is the mass to luminosity ratio of the population,
    $M_{gc}$ is the mean GC mass ($2.4 \times 10^{5}$ $M_\odot$), and
    the numerical 
    factor c is the luminosity corresponding to $M_{\lambda}=-15.0$
    in solar luminosity units.

    Adopting the specific frequencies derived in Section 7, the mass
    to luminosity ratios discussed in the previous section, with
    $c=1.55 \times 10^{8}$ for the blue band, yield
    $\eta_{RGC}=1.25 \times 10^{-3}$ and $\eta_{BGC}=7.60 \times 10^{-3}$.

    Beside, and following \citealt{McL99}, this definition 
    can be revised in terms of the total baryonic mass, i.e., including the 
    mass of hot gas. 
    In this case we define a shell with inner and outer (spatial) radii 
    of 11 and 66 Kpc respectively. The adoption of the inner boundary
    attemps to decrease the effect of gravitational disruption on
    the GC sample while the outer one is the limit of our brightness
    profile. About 1800 RGC and 1500 BGC fall within these boundaries
    that enclose $3.54 \times 10^{11}$ and $0.55 \times 10^{11}$ $M_\odot$
    of stars of the red and blue populations, respectively.
    On the other side, adopting the X ray analysis by Jones et al. (1997)
    and correcting their adopted distance (24 Mpc) to ours, we obtain 
    a hot gas mass of $0.28\times 10^{11} M_\odot$. These quantities lead
    to an overall efficiency $\eta=1.8 \times 10^{-3}$ for both GC populations
    combined. This value is somewhat smaller, but compares well, with the GC 
    formation efficiency ($2.9 \times 10^{-3}$) derived by \citet{McL99}.

\section{A comparison Between Globular Clusters, Stars, Hot Gas and Dark
        Matter Distributions}
\label{COMP}

\subsection{Red Globular Clusters and Total Galaxy Mass}

        The results presented in the previous sections are consistent with
        the existence of a link between the RGC and the (red and dominant) 
        halo stellar population of NGC 1399. This connection is also
        supported by the GC kinematic behaviour. For example, \citet{Sag2000}
        give a radial velocity dispersion for the halo stars of about
        280 $km~s^{-1}$ (at 120 arcsec from the centre) in good agreement
        with the RGC value of 265 $km~s^{-1}$ (from R2004 data, without
        the velocity cut-off adopted by these authors). Taking into account
        that the derived volumetric profile for these clusters is consistent
        with a logarithmic spatial density derivative $(dln(\rho)/dlnr)=-2.9$
        (at galactocentric radii larger than $\approx 100$ arcsec), and using
        the Jeans equation for the spherical and isothermal case, the resulting
        total mass within 100 Kpc of galactocentric radius ranges from
         $4.4$ to $4.8\times 10^{11} M_\odot $ (for an anisotropy parameter $\beta$ 
         from $1.0$ to $-0.5$).

\subsection{Blue Globular Clusters, Hot Gas and Dark Matter}

\begin{figure}
\includegraphics[width=9cm,height=9cm] {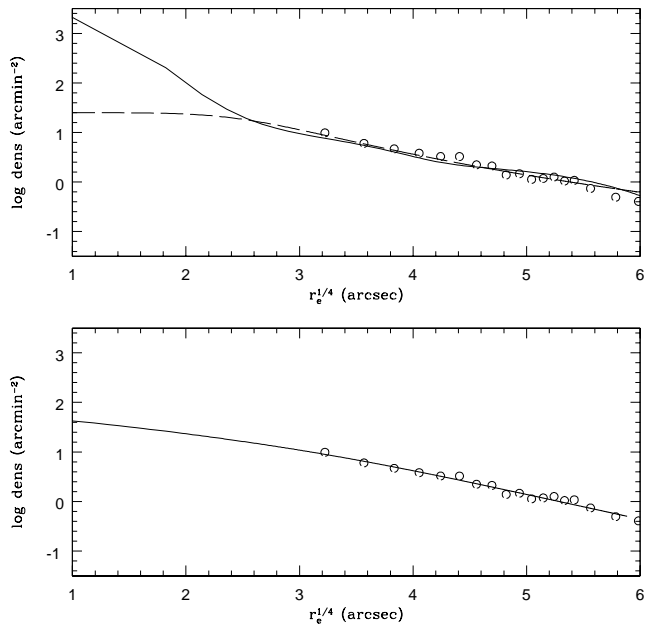}
\caption{Upper panel: The NGC 1399 blue globular clusters density profile (open circles)
       compared with that of the hot X ray emitting gas (continuous line: from
       \citealt{Pao2002}; dahed line: from \citealt{Jones97} . Lower panel:
       blue GC compared with the (projected) dark matter profile obtained
       by R2004 using GC kinematics.}
\label{DMGC} 
\end{figure}

        The X ray emission of the Fornax cluster has been the subject
        of several works (e.g. \citealt{Serle93}; \citealt{Ikebe92}). 
        ROSAT observations have been discussed by Jones et al.
        (1997) who fit a surface brightness profile of the form:

\begin{eqnarray}
S(r)=S(0) \left[1+\left( \frac{r}{a} \right) ^2 \right]^{-3\beta+0.5}
\end{eqnarray}
\noindent
         centered on NGC 1399, obtaining best fitting parameters $a=45$ arcsec 
         and $\beta=0.35$. In turn, \citet{Pao2002} give a thorough
         description of the complex X ray emission and suggest a composite
         beta model (including ``central'', ``galaxy'' and ``cluster''
         components). Both the \citet{Jones97} and \citet{Pao2002} profiles
         are depicted in Figure~\ref{DMGC} (upper panel) and exhibit a good coincidence
         with the surface density of the blue GC.

         The lower panel in Figure 13 shows the projected profile that gives a 
         representation of the dark matter distribution (also shifted in 
         ordinates) derived by R2004 on the basis of their GC kinematic study. 
         These last authors note that a NFW profile 
         (i.e. $\zeta=1.0$; \citealt{NFW96}; \citealt*{NFW97}) with a scale length 
         $r_s=394$ arccsec gives a good representation of the inferred 
         volumetric density of dark matter. We remark that, although their fit 
         covers a galactocentric range from approximately 2 to 8 arcmin 
         (${r_e}^{1/4}$ from 3.31 to 4.68), the profile extrapolation holds up 
         to 20 arcmin (i.e. the approximate limit of the D2003 data). For this
         reason, and among all the acceptable sets of $\zeta-rs$ values, we
         adopted $\zeta=1.0$ and $r_{s}=375$ arcsec as adequate for the
         BGC noting that our scale length can be considered as identical
         (to within the uncertainties) with the value given in R2004.
         
         As discussed in Section~\ref{VDP}, volumetric profiles with different
         values of the $\zeta$ parameter can yield almost equivalent fits 
         to the observed areal densities for galactocentric radii larger 
         than 2 arcmin. This means that, beyond the very good agreement
         between the distribution of the blue GC and dark matter, the
         data do not allow us to support unambiguously that an NFW profile
         also holds in the inner regions of the galaxy. However, all these
         profiles are consistent with an spatial derivative
         $dln(\rho)/dln(r) \approx -1.4$ from 100 arcsec and outwards in
         galactocentric radius. If this value is adopted as representaive
         for the hot gas (according to Figure 13) and for a temperature
         range of 1.1 to 1.4 Kev (Jones at al. 1997; Paolillo et al. 2002), the total gravitating mass within
         100 Kpc assumig hydrostatic gas equilibrium, becomes
          $5.3$ to $5.8\times 10^{11} M_\odot$, i.e., comparable with that derived from
          the RGC kinematics mentioned before.

         The fact that the slope of the density profiles of the blue GC
         and that corresponding to the X ray emission associated with
         NGC 1399 seem similar was already suggested by \citet{HIR99}.
         In turn, the similarity between the distribution of the blue GC and
         that of dark matter was also pointed out by \citet{CMW98} in
         their analysis of the \textit{outer} regions of the M87 GCS.

\section{Discussion}
\label{DISC}

     The surface brightness profiles presented in Section~\ref{BandR1399}
     cannot be matched with a single $r^{1/4}$ law, a feature already noted
     in previous works (e.g. \citealt{Scho86}) that have also pointed out
     the presence of an outer corona that would entitle the classification
     of NGC 1399 as a cD galaxy (but see the difficulties inherent
     to such a classification described in R2004).

     If in fact the formation of a given GC family occurs concomitantly
     with that of a diffuse stellar population that shares its age and
     chemical abundance, it may be expected that, after gravitational
     relaxation, both stars and clusters will exhibit similar spatial
     distributions. Although dynamical friction might play a differential
     role that remains to be estimated, globulars could be identified as
     tracers of both the spatial distribution and of the kinematic behaviour
     of coeval diffuse stellar populations. Then, the galaxy brightness profile
     would be the result of the ``luminosity weighted'' composition of
     those events. This approach gives a proper fit of the NGC 1399
     profile shape by assuming the presence of
     two dominant populations (although they might exhibit their own degree
     of internal inhomogeneity in terms of age and chemical abundance)
     typified by very different intrinsic specific frequencies $S_n$.
     As the blue stellar population becomes more prominent outwards,
     due to its shallower areal density slope, the degree of "cDness" of
     the galaxy can also be identifyied as a result of that population.

     An important result of the analysis is that, even though both the red 
     and blue GC families do not exhibit a detectable change of their
     modal colours over a galactocentric range close to 100 Kpc, the
     composition of their associated diffuse stellar populations do lead 
     to a blueward colour gradient that compares well with the observations.

     A caveat remains, however, in the sense that the galaxy profile
     can be matched by adopting a stellar population density profile
     similar to that of the red GC and forcing a spatial colour
     gradient. This would imply, on the one hand, that for some reason
     the stellar and red GC enrichments were decoupled and, on the other,
     that blue GC formed in a process unable to form non-clustered stars.

     Although the overall globular cluster formation efficiency obtained
     in previous sections seems comparable to other values available
     in the literature (e.g. \citealt{McL99}) the considerably higher
     intrinsic $S_n$ value of the blue population in comparison with the red one
     also implies a higher globular formation efficiency that may be
     indicative of a distinct formation environment.
     Possible explanations of this situation have been given in the
     literature in connection with the role of re-ionization as
     responsible of the termination of the star forming process or
     in the frame of the \citet{SZ} scenario. In this last case, 
     relatively small gas clumps associated with the formation of the BGC are
     disrupted by Supernovae.

     As noted before, the profile fit does not give an acceptable
     description of the inner 30 arcsec in galactocentric radius where
     the galaxy is brighter and redder than predicted. This situation 
     may arise from the use of an inappropriate analytical profile,
     and/or the effect of GC disruption by gravitational forces
     (see, for example, \citealt{Vespe2003}). Aperture
     integrated $(C-T_1)$ colours measured by \citet{OFG98}
     are indicative of the presence of a high metallicity stellar
     population (\citealt{Pick85}; \citealt{Kun2000}). This situation has
     similarities with other galaxies (e.g. the MW; \citealt{McWR}) where
     bulge stars seem to have metallicities somewhat higher than those
     of the RGC.
     A distinctive kinematic behavior
     of the NGC 1399 central region can be seen in the velocity dispersion profile
     obtained by \citet{Sag2000}. These facts indicate that the galaxy
     nucleus may have harbored a different sequence of stellar formation
     events disparate from    those observed over much larger spatial scale
     lengths.

\section{Conclusions: A tentative Scenario}
\label{CONCL}

      Previous sections have described a possible approach to the
      link between GC and the diffuse stellar population in
      NGC 1399 on a purely photometric basis. Although spectroscopic
      results \citep{Forb2001} indicate the existence of an
      age spectrum among its GC (and, presumably, in the associated
      diffuse stellar populations), the analysis of a significant
      volume of spectroscopic data in terms of age and metallicity is
      still missing.

      In a cautious way, however, the results presented in the previous 
      sections, and other available in the literature, can be ordered into
      a working scenario that attempts to delineate a temporal sequence
      of events that might end in the observed situation, namely:
\begin{description}

\item [-] The blue GC in NGC 1399 exhibit modal integrated colours that 
      are comparable to those observed in other galaxies and are consistent
      with a frequently mentioned fact: The relative homogeneity of the blue
      GC population from one system to another, suggesting that
      their origin might have occurred before the formation of the dominant
      stellar population of the galaxy haloes.
      However, recent works (\citealt{LOTZ}; \citealt{STRA}) indicate that a
      correlation between the integrated BGC colors and the galaxy
      absolute magnitude may exist. These results imply that, even though the
      BGC might have formed in a very early stage, these clusters "knew" about
      the galaxy they were forming in.

\item [-] There is a remarkable coincidence of the spatial distribution of
      the blue GC with that inferred for dark matter by R2004 on the
      basis of their kinematic study. We stress that, even though these
      last authors perform their analysis in a galactocentric range
      from 2 to 8 arcmin, the extrapolation of their dark mass profile
      (projected on the sky) matches that of the BGCs up to 20 arcmin
      (the outer limit in D2003 and in our analysis).

\item [-] X ray emitting hot gas, usually assumed to be at the virial 
      temperature of galaxy clusters, also show a spatial distribution 
      that, within the uncertainties, is coincindent with that of the
      blue GC. We note that the mean metallicity of this gas ($[Fe/H]=-0.4$;
      \citealt{Jones97}) is comparable to that of the RGC.

\item [-] A revision of the R2004 kinematic data suggests  that at least the
outer blue GC
      in NGC 1399 may exhibit, as in other galaxies, a detectable amount
      of ``solid-body''-like rotation. The presence of angular momentum
      in these clusters has been noted, for example, in M87 \citep{KPG98} and NGC 4472 \citep{CMCB2003}.

\end{description}

      All these features strongly suggest that blue GC may have had their
      origin at a very early phase of stellar formation, sharing the
      evolution of dark matter clumps in the non dissipative collapse
      that finally formed the most massive component of galaxy haloes,
      a fact that might also be reflected by their high radial velocity 
      dispersion. In this context, the usefulness of simple closed box-type
      chemical evolution models has been noted, for example, by \citet{HH2000},       \citet{DHP} and, in the particular case of the MW,
      by \citet{VDH}. Along these lines, the scenario described by
      Harris \& Harris (2002) for NGC 5128, where early star formation
      goes on with an initial stage of rapid infall of metal poor gas,
      and based on the similarities with NGC 1399 noted in Section 9,
      may also seem relevant for this last galaxy.

      In turn, red GC seem connected with the formation of the dominant
      stellar population, probably at the expense of gas that has
      suffered a rather inhomogeneous dissipative collapse leading to
      angular momentum loss and to their broad chemical abundance
      distribution. This process might have been boosted by the effect
      of the very early enrichment provided by the blue GC (and the
      associated diffuse stellar population).

      The idea that GC may have played a role as early ``seeds'' of chemical
      enrichment was already put forward by \citet*{YLS83}.
      The main problem with their model is that the small mass locked
      in GC could not provide enough heavy elements to increase the
      overal metallicity level. This  problem is overcome if, in fact,
      most of the enrichment comes not from the cluster themselves but
      from a much more massive (although less conspicuous) diffuse stellar
      population. The role of the oldest GC and its associated stellar 
      population in the re-ionization of the
      Universe would also be enhanced on the same argument.

      As a reference, we note that such a blue stellar population might
      provide some $1.9\times10^9M_\odot$ of heavy elements within
      a galactocentric radius of 100 Kpc. This mass of heavy elements
      arises in an estimated stellar mass of $8.1\times10^{10} M_\odot$
      contained in the blue population (from the blue stellar light profile 
      and the (M/L) ratios given in Section 9) and adopting the net yields
      ($\eta=0.023$) given by \citet{PORT} and revised by Cora (2004 in
      prep.) for a very low metallicity population. If the current red
      stellar population ($8.1\times10^{10} M_\odot$) was mostly in gaseous
      form at that stage, the injected heavy metals would have rised the
      chemical abundance to about $0.15Z_\odot$.

      This enrichment
      is about one third of the mean abundance of the red stellar
      population (inferred from the associated RGC), i.e, not
      enough to explain the observed differences between both populations.
      However, that level of enrichment might have had an important
      impact on the gas cooling function, improving the star formation
      efficiency and triggering the process that ended up with the
      broad abundance distribution of the red stellar population
      (and also on the current chemical composition of the hot gas).

      The much higher formation efficiency of the blue clusters suggest
      very different environmental conditions from those that prevailed
      for the red globulars. This fact, and the impact of a diffuse
      and low metallicity  stellar population on the subsequent process
      of galaxy formation remains worthy of exploration.

      A final comment about the possible existence of such a
      stellar population around NGC 1399 deals
      with the fact that, being that galaxy in the neighborhood of
      the Fornax cluster barycenter, it is not clear if the association is 
      with this galaxy or with the cluster as a whole. For example, 
      \citep{BCFD} argue that GC, apparently not associated with galaxies
      in the field, can be identified at angular distances as large
      as 4800 arcsecs from NGC 1399 (a distance comparable to the
      galaxy virial radius); the eventually associated diffuse population
      at that radius would be hard to detect as the expected blue surface
      brightness would be in the range from 31 or 32 mag per sq. arcsec.

\section*{Acknowledgments} 
      The authors are grateful to Dr. Juergen Scheer for allowing
      the use of his data. One of us (FF) carried out this work with a       
      Research Fellowship from CONICET . This work was funded
      with grants from CONICET, ANPCYT and Universidad Nac. de La Plata
      (Argentina).


\clearpage

\begin{thebibliography}{}
\bibitem[\protect\citeauthoryear{Ashman, Conti \& Zepf}{Ashman et al.}{1995}]{ACZ95} Ashman K.M., Conti A., Zepf S.E., 1995, AJ, 110, 1164
\bibitem[\protect\citeauthoryear{Bassino et al.}{2003}]{BCFD} Bassino L., Cellone S.A., Forte J.C., Dirsch B., 2003, A\&A, 399, 489
\bibitem[\protect\citeauthoryear{Baugh et al.}{1998}]{BAUGH} Baugh C. M., Cole S., Frenk C. S., Lacey C. G., 1998, ApJ, 498, 504
\bibitem[\protect\citeauthoryear{Beasley et al.}{2002}]{B2002} Beasley M.A., Baugh C.M., Forbes D., Sharples R.M., Frenk C.S.,2002, MNRAS, 347, 1150
\bibitem[\protect\citeauthoryear{Binney \& Tremaine}{1987}]{BT87} Binney J., Tremaine S., 1987, in Galactic Dynamics, Princeton Univ. Press
\bibitem[\protect\citeauthoryear{Caon et al.}{1994}]{CCD94} Caon N., Capaccioli M., Donofrio M., 1994, A\&AS, 106, 199
\bibitem[\protect\citeauthoryear{Canterna}{1976}]{C76}Canterna R. 1976, AJ, 81, 228
\bibitem[\protect\citeauthoryear{Capuzzo-Dolcetta \& Tesseri}{1999}]{CDT99} Capuzzo-Dolcetta R., Tesseri A., 1999, MNRAS, 308, 961
\bibitem[\protect\citeauthoryear{Capuzzo-Dolcetta \& Donnarumma}{2001}]{CDD2001} Capuzzo-Dolcetta R., Donnarumma, I., 2001, MNRAS, 328, 645
\bibitem[\protect\citeauthoryear{Cohen et al.}{1998}]{CBR98} Cohen J. G., Blakeslee J.P., Rhyzov A., 1998, ApJ, 496, 808
\bibitem[\protect\citeauthoryear{C\^ot\'e et al.}{1998}]{CMW98} C\^ot\'e P., Marzke R.O., West M. J., 1998, ApJ, 501, 554
\bibitem[\protect\citeauthoryear{C\^ot\'e et al.}{2003}]{CMCB2003} C\^ot\'e P., McLaughlin D.E., Cohen J.G., Blakeslee J.P. 2003, ApJ, 591, 850
\bibitem[\protect\citeauthoryear{De Young, Lind \& Strom}{De Young et al.}{1983}]{YLS83} De Young D. S.,Lind K., Strom S.E. 1983, PASP, 95, 401
\bibitem[\protect\citeauthoryear{Dirch et al.}{2003}]{D2003} Dirsch B., Richtler T., Geisler D., Bassino L.P., Gieren W.P., 2003, AJ, 125, 1908
\bibitem[\protect\citeauthoryear{Durrell, Harris \& Pritchet}{2001}]{DHP} Durrelll P. R., Harris E. W., Pritchet C. J., 2001, AJ, 121, 2557
\bibitem[\protect\citeauthoryear{Eggen, Lynden Bell \& Sandage}{Eggen et al.}{1962}]{ELS62} Eggen O., Lynden Bell D., Sandage A., 1962, ApJ, 136, 748
\bibitem[\protect\citeauthoryear{Forbes et al.}{1997}]{FBH97} Forbes D.A., Brodie J.P., Huchra J., 1997, AJ, 113, 887
\bibitem[\protect\citeauthoryear{Forbes et al.}{1998}]{FBG98} Forbes D.A., Brodie J.P., Grillmair C.J., 1998, MNRAS, 293, 325
\bibitem[\protect\citeauthoryear{Forbes et al.}{2001}]{Forb2001} Forbes D.A., Beasley M.A., Brodie J.P., Kissler-Patig M., 2001, ApJ, 563, 143
\bibitem[\protect\citeauthoryear{Forbes \& Forte}{2001}]{FF2001} Forbes D.A., Forte J.C, 2001, MNRAS, 322, 257
\bibitem[\protect\citeauthoryear{Forte, Strom \& Strom}{Forte et al.}{1981}]{FSS81} Forte J.C., Strom S.E., Strom K., 1981, ApJ, 245, L9
\bibitem[\protect\citeauthoryear{Forte et al.}{2002}]{Forte2002} Forte J.C., Geisler D., Kim E., Lee M.G., Ostrov P.G., 2002 in Geisler D., Grebel E.K. \& Minniti D.,eds., Proceedings of the
 207th IAU Symposium, Extragalactic Star Clusters, Astron. Soc. Pac., San Francisco, p. 251
\bibitem[\protect\citeauthoryear{Geisler}{1996}]{Geisler96} Geisler D., 1996, AJ, 111, 480
\bibitem[\protect\citeauthoryear{Geisler \& Forte}{1990}]{GF90} Geisler D., Forte J.C., 1990, ApJ, 350, L5
\bibitem[\protect\citeauthoryear{Geisler, Lee \& Kim}{Geisler et al.}{1996}]{GLK96} Geisler D., Lee, M. G., Kim E. , 1996, AJ, 111, 1529
\bibitem[\protect\citeauthoryear{Grillmair et al.}{1999}]{GFBE99} Grillmair C.J., Forbes D.A., Brodie J.P., Elson R.A.W., 1999, AJ, 117, 167
\bibitem[\protect\citeauthoryear{Harris et al.}{1977}] {HC77} Harris H., Canterna R., 1977, AJ, 82, 798
\bibitem[\protect\citeauthoryear{Harris}{1996}]{Harris96} Harris W.E., 1996, AJ, 112, 1487
\bibitem[\protect\citeauthoryear{Harris \& van der Bergh}{1981}]{HB81} Harris W.E., van den Bergh S., 1981, AJ, 86, 1627
\bibitem[\protect\citeauthoryear{Harris, Harris \& McLaughlin }{Harris et al.}{1998}]{HHM98} Harris W.E., Harris G.L.H, McLaughlin D.E., 1998, AJ, 115, 1801 
\bibitem[\protect\citeauthoryear{Harris \& Harris}{2000}]{HH2000}  Harris G. L. H., Harris W. E., 2000, AJ, 120, 2423
\bibitem[\protect\citeauthoryear{Harris \& Harris}{2002}]{HH2002} Harris W.E., Harris G.L.H., 2002, AJ, 123, 3108
\bibitem[\protect\citeauthoryear{Harris}{2003}]{Harris3} Harris, W. E. 2003, Extragalactic Globular Cluster Systems, Proceedings of the ESO Workshop held in Garching, Germany, Agust 27-30  2002, p. 317.
\bibitem[\protect\citeauthoryear{Hilker et al.}{1999}]{HIR99} Hilker M., Infante L., Richtler T., 1999, A\&AS, 138,55
\bibitem[\protect\citeauthoryear{Ikebe et al.}{1992}]{Ikebe92} Ikebe Y. et al. 1992, ApJ, 384, L5
\bibitem[\protect\citeauthoryear{Jones et al.}{1997}] {Jones97} Jones C., Stern C., Forman W., Breen J., David L., Tucker W., 1997, ApJ, 482, 143
\bibitem[\protect\citeauthoryear{Killeen \& Bicknell}{1988}]{KB88}    Killeen N.E.B., Bicknell G.V., 1988, ApJ, 325, 165
\bibitem[\protect\citeauthoryear{Kissler-Patig \& Gebhardt}{1998}]{KPG98} Kissler-Patig M., Gebhardt K. 1998, AJ, 116, 2237
\bibitem[\protect\citeauthoryear{Kissler-Patig}{2002}]{KP2002} Kissler-Patig, M. 2002, in Geisler, D., Grebel E.K, \& Minniti D., eds., Proceedings of the IAU 207th Symp., 
 Extragalactic Star Clusters, Astron. Soc. Pac., San Francisco, p. 207
\bibitem[\protect\citeauthoryear{Kissler-Patig}{2003}] {KP2003} Kissler-Patig, M. 2003, Extragalactic Globular Cluster Systems, Proceedings of the ESO Workshop held in Garching, Germany, Agust 27-30  2002, ESA.
\bibitem[\protect\citeauthoryear{Kundu \& Whitmore}{1998}]{KW98}  Kundu A., Whitmore, B.C., 1998, AJ, 116, 2841
\bibitem[\protect\citeauthoryear{Kuntschner}{2000}]{Kun2000} Kuntschner H., 2000, MNRAS, 315, 184
\bibitem[\protect\citeauthoryear{Larsen \& Richtler}{2000}]{LR2000} Larsen S. S., Richtler T., 2000, A\&A, 354, 836L
\bibitem[\protect\citeauthoryear{Lee, Lee \&  Gibson}{lee et al.}{2002}]{LLG2002} Lee H-c, Lee Y.W., Gibson B.K., 2002, AJ, 124, 2664
\bibitem[\protect\citeauthoryear{Lotz, Miller \& Ferguson}{2004}]{LOTZ} Lotz J. M., Miller B. W., Ferguson H. C., 2004, AJ, 613, 262
\bibitem[\protect\citeauthoryear{McLaughlin}{1999}]{McL99} McLaughlin D.E., 1999, AJ, 117, 2398
\bibitem[\protect\citeauthoryear{McWilliam \& Rich}{1994}]{McWR} McWilliam, A., Rich, R. M., 1994, ApJS, 91, 749
\bibitem[\protect\citeauthoryear{Michard}{2000}]{Mich2000} Michard R., 2000, A\&A, 360, 85
\bibitem[\protect\citeauthoryear{Navarro et al.}{1996}]{NFW96} Navarro J.F., Frenk C.S., White S.D.M., 1996, ApJ, 462, 563
\bibitem[\protect\citeauthoryear{Navarro, Frenk \& White }{Navarro et al.}{1997}]{NFW97} Navarro J.F., Frenk C.S., White S.D.M., 1997, ApJ, 490, 493
\bibitem[\protect\citeauthoryear{Ostrov, Geisler \& Forte}{1993}]{OGF93} Ostrov P.G., Geisler D., Forte J.C., 1993, AJ, 105, 1762
\bibitem[\protect\citeauthoryear{Ostrov, Forte \& Geisler}{Ostrov et al.}{1998}]{OFG98} Ostrov P.G., Forte J.C., Geisler D., 1998, AJ, 116, 2854
\bibitem[\protect\citeauthoryear{Paolillo et al.}{2002}]{Pao2002} Paolillo M., Fabbiano G., Peres G., Kim D.-W. 2002, ApJ, 565, 883
\bibitem[\protect\citeauthoryear{Peebles \& Dicke}{1968}]{PD68}  Peebles P.J.E., Dicke R.H. 1968, ApJ, 154, 891
\bibitem[\protect\citeauthoryear{Pickles}{1985}]{Pick85} Pickles A. J., 1985, ApJ, 296, 340
\bibitem[\protect\citeauthoryear{Portinari, Chiosi \& Bressan}{1998}]{PORT} Portinari L., Chiosi C., Bressan A., 1998, A\&A, 334, 505
\bibitem[\protect\citeauthoryear{Reed, Hesser \& Shawl}{Reed et al.}{1988}]{RHS88} Reed B.C., Hesser J.E., Shawl S.J., 1988, PASP, 100, 545
\bibitem[\protect\citeauthoryear{Richtler et al.}{2004}]{R2004} Richtler T., Dirsch B., Geisler D., Gebhardt K. Hilker M., Alonso M.V., Forte J.C., Grebel E.K., Infante L., Larsen S., Minniti D., Rejkuba M., 2004, AJ, 127, 2094
\bibitem[\protect\citeauthoryear{Ricotti}{2002}] {Rico2002} Ricotti, M. 2002, MNRAS, 336, L33
\bibitem[\protect\citeauthoryear{Saglia et al.}{2000}]{Sag2000} Saglia R.P., Kronawitter, A., Gerhard, O., Bender, R., 2000, AJ, 119, 153
\bibitem[\protect\citeauthoryear{Santos}{2003}]{San2003} Santos M.R. 2003, in Extragalactic Globular Cluster Systems, Proceedings of the ESO Workshop held in Germany, 27-30 Aug. 2002, p. 348
\bibitem[\protect\citeauthoryear{Schlegel et al.}{1998}]{SFD98} Schlegel D., Finkbeiner D., Davis M., 1998, ApJ, 500, 525
\bibitem[\protect\citeauthoryear{Scheer}{1987}]{Scheer87} Scheer J. 1987, Appl. Optics, 26, 3077
\bibitem[\protect\citeauthoryear{Schombert}{1986}]{Scho86} Schombert J.M., 1986, ApJS, 60, 603
\bibitem[\protect\citeauthoryear{Schweizer}{1987}]{Schw87} Schweizer F., 1987, Nearby Normal Galaxies. From the Planck Time to the Present. Springer, New York, p. 18.
\bibitem[\protect\citeauthoryear{Searle \& Zinn}{1978}]{SZ} Searle L., Zinn R., 1978, ApJ, 205, 357
\bibitem[\protect\citeauthoryear{Serlemitsos et al.}{1993}]{Serle93} Serlemitsos P., Loewenstein M., Mushotzky R., Marshall F., Petre R., 1993, ApJ, 413, 518
\bibitem[\protect\citeauthoryear{Strader, Brodie \& Forbes}{2004}]{STRA} Strader J., Brodie J. P., Forbes D. A., 2004, AJ, 127, 3431
\bibitem[\protect\citeauthoryear{VanDalfsen \& Harris}{2004}]{VDH} VanDalfsen M. L., Harris E. W., 2004, AJ, 127, 368
\bibitem[\protect\citeauthoryear{Vesperini et al.}{2003}]{Vespe2003} Vesperini E., Zepf S.E., Kundu A., Ashman K.M., 2003, ApJ, 593, 760 
\bibitem[\protect\citeauthoryear{Worthey}{1994}]{Wor94} Worthey G., 1994, ApJS, 95, 107
\end{thebibliography}
\end{document}